\documentclass[12pt,preprint]{aastex}
\usepackage{emulateapj5}

\shorttitle{Evidence for Bare Strange Stars?}

\shortauthors{R.X. Xu}

\begin{document}

%\slugcomment{Submitted to ApJ Part 1}

\title{Thermal Featureless Spectrum: Evidence for Bare Strange Stars?}

\author{R.X. Xu}
\affil{School of physics, Peking University, Beijing 100871,
China; rxxu@bac.bku.edu.cn}

\begin{abstract}

Motivated by trying to understand the absence of spectral lines in
the thermal components of the X-ray compact sources observed
recently by Chandra or XMM, we propose that these sources could be
simply bare strange stars.
The formation, cooling, and thermal photon radiation of bare
strange stars have been investigated.
It is suggested that thermal featureless spectrum could be a new
probe for identifying strange stars.


\end{abstract}

\section{Introduction}

To affirm or negate the existence of strange star is an exciting
and meaningful approach to guiding physicists in studying the
quantum chromodynamical (QCD) nature of strong interaction.
With regard to those three possible ways of finding strange stars
(see, e.g., Xu \& Busse 2001, for a short review), a {\em hard}
evidence to identify a strange star might be found by studying
only the surface conditions since the other two avenues are
subject to many complex nuclear and/or particle physics processes
poorly known.
New advanced X-ray detectors, the Chandra and the XMM, increase
the possibility of discovering the surface differences between
neutron star (NS) and bare strange star (BSS) since both exteriors
of NS and BSS should be thermal X-ray radiators.

Many calculations, first developed by Romani (1987) and then by
others (e.g., Zavlin et al. 1996), show that spectral lines form
in the atmospheres of NS or crusted strange stars (CSS), which
should be detectable with the spectrographs on board Chandra and
XMM.
However {\it none} of the sources reported recently has
significant spectral features in the observations with Chandra or
XMM. These sources are collected in Table 1, the spectra of which
can be well fitted with blackbody models for the thermal
components.
An observation presented by Marshall \& Schulz (2002) indicates
still no significant line feature even in the pulse-phased spectra
over the 0.15-0.80 keV band, which does not favor both models of
atmospheres of heavy elements or pure hydrogen.
Although this discrepancy could be explained for some of the
sources by assuming a low-Z element (hydrogen or helium)
photosphere or by adjusting the magnetic field, a simple and
intuitive suggestion, which will be addressed in this {\em
Letter}, for the explanation is that these ``neutron stars'' are
actually just BSSs, especially the nearest one RX J1856.5-3754 (no
NS atmosphere model available can fit its X-ray {\it and} optical
data, Burwitz et al. 2001), since almost no atom appears above a
bare quark surface.
%
%Burwitz et al. (2001): ``Detailed spectra analysis of the combined
%X-ray and optical data rules out the nonmagnetic neutron star
%atmosphere models with hydrogen, helium, iron and solar
%compositions. We also conclude that strong magnetized atmosphere
%models are unable to present the data.'' --- to be the nearest NS?
%(see Heiseberg 2002 for the answer)

Strange stars can be bare and BSSs can exist as compact stars (Xu
\& Qiao 1998, Xu, Zhang \& Qiao 2001).
Drifting subpulses of radio pulsars could be an evidence for BSSs
(Xu, Qiao \& Zhang 1999), which represent the strong binding of
particles above pulsar surfaces.
Super-Eddington emission of soft $\gamma-$ray repeaters
%(the giant flares of SGR 0526-66 on 1979/03/05 have peak
%luminosity $\sim 10^{45}$ erg s$^{-1}$, 7 orders of magnitude in
%excess of the Eddington limit)
might be another evidence for BSSs (Zhang, Xu \& Qiao 2001, Usov
2001a).
If further observations with much higher signal-to-noise ratios
still find no spectral feature in the thermal components of the
sources listed in Table 1, the featureless thermal spectrum should
be a new evidence for BSSs.

%\section{Can bare strange stars exist in nature?}
\section{Strange stars: crusted or bare?}
%{Accretion scenarios in pulsars life}

Current researches show that it is possible to leave a strange
star after a core-collapse type supernova explosion. However, it
depends on wether the strange star is bare or crusted to
distinguish neutron stars and strange stars based on their sharp
differences in surface conditions.
Although relevant qualitative arguments are addressed separately
in many published papers, it is worth summarizing concisely those
discussions, with the inclusion of some quantitative estimates and
modifications.
A protostrange star may be bare due to strong mass ejection and
high temperature (Usov 1998), whereas a BSS may still have the
possibility to be covered by a crust due to the accretion of (1)
supernova fallback and of (2) debris disk. It follows below that
such an accretion-formed crust looks probably in no likelihood.

In case (1), as Xu et al. (2001) suggest, due to rapid rotation
and strong magnetic field, most of the fallback matter may form
temporarily a fossil disk,
%the implication of which has been widely discussed in explaining
%the properties of the anomalous X-ray pulsars (AXPs) and soft
%$\gamma$-ray repeaters (SGRs) (e.g., Marsden \& White 2001),
and the initial accretion onto a star is almost impossible.
Recently, 3-D simulations by Igumenshchev \& Narayan (2002) show
that the gravitational energy of the infall magnetized plasma has
to be converted to other energies, and that the initial accretion
rate could be reduced significantly.
%
%Nevertheless one can still estimate the accretion rate in this
%case by the dimensional argument in Xu et al. (2001), which could
%be $\Delta M\sqrt{2GM}/r_{\rm c}^{3/2}$, where $\Delta M$
%presented in Eq.(5) of Xu et al. (2001) is the total mass of
%fallback material trapped by the the magnetosphere, $r_{\rm
%c}=1.5\times 10^8 M_1^{1/3}P^{2/3}$ cm is the co-rotating radius,
%$P$ the rotation period.
%
%This accretion energy rate ${\cal L}_1\sim$ ergs/s.
%
Nevertheless, Xu et el. (2001) proposed another trapping of
supernova ejecta by magnetic fields, rather than the Chevalier's
(1989) one by gravity. This trapped material with mass $\sim
10^{-15}M_\odot$ could fall back onto the surface to form a
massive atmosphere {\em if} there exits no other force but
gravity. However, the radiative pressures of {\em strong} photon
and neutrino emission are not negligible because of high
temperature ($T\sim 10^{11}-10^8$ K, see \S3). Only more than
years latter could the photon luminosity be smaller than the
Eddington one ($L_{\rm Edd}\sim 10^{38}$ ergs/s). A possible
scenario could be as follow. The trapped ions, forced by
radiation, move along the magnetic lines to an out most region of
each line, where these enriched ions go across the field lines
(higher density increases the kinematic energy density), and may
merge eventually into debris disk.
%Note that the inner most radius of the disk is likely lager than
%the co-rotating radius.

Accretion in case (2) was supposed to power the X-ray emission of
AXPs and SGRs (e.g., Marsden \& White 2001) during the
non-stationary ``prospellar'' phase, since the X-ray powers of
these sources are much higher than the energy loss rates of their
spindowns.
These accretion energy rates ${\cal L}$ could be expected ${\cal
L}<10^{36}$ ergs/s (the maximum persistent X-ray luminosity of
AXPs and SGRs).
Can a crust be formed during such accretions?
Because of strong fields, infalling matter is funneled toward the
polar caps, goes like freefall until feeling the deceleration due
to the radiation pressure generated by the accreted material on
the caps. Without this halting, a proton could have kinematic
energy of $\sim 100$ MeV near the surface of a BSS, and could thus
penetrate the Coulomb barrier ($\sim 20$ MeV) and dissolves. But
in a radiation field with energy density $U$, a hydrogen atom will
actually have a back force $f_{\rm r}\sim \sigma_{\rm T}U$ with
$\sigma_{\rm T}$ the Thompson cross section. For low accretion
limit, such force is not negligible only when atoms are near the
hot spot powered by accretion, and the height of this region is
about the polar cap radius, $r_{\rm p}$. In view of only
$\varepsilon$ times of the accretion energy has been re-emitted
above polar cap (see Appendix), by $f_{\rm r}r_{\rm p}=GMm_{\rm p
}/R$, we obtain a critical accretion rate ${\cal L}^*$,
\begin{equation}
{\cal L^*}={2\pi G c r_{\rm p}Mm_{\rm p}\over
\varepsilon\sigma_{\rm T}R}={9.1\times 10^{35}\over
\varepsilon}M_1R_6^{-1}P^{-1/2}~~{\rm ergs~s^{-1}}, \label{L*}
\end{equation}
which is $\sim 1/\varepsilon$ times higher than the critical value
presented by Basko \& Syunyaev (1976).
If ${\cal L}<{\cal L^*}$, an atom may still have enough kinematic
energy to penetrate after the deceleration. We expect that the
accretion of ISM or fossil disk can also keep a strange star to be
bare since ${\cal L}<10^{36}$ ergs/s $<{\cal L^*}$.
It is very likely that ${\cal L}^*>L_{\rm Edd}$, since
$\varepsilon$ can be as small as $10^{-4}$ (see Appendix). This
means that BSSs may also survive some of the accretions with
super-Eddington rates.
A crust covering a strange star could be formed via accretion if
${\cal L}>{\cal L^*}$, which ensures no distinction between the
faced and magnetospheric radiations of SSs and NSs.
But such a high accretion rate might be only possible for binary
X-ray sources.
Recycled millisecond pulsars could be BSSs as long as the
accretion rates $<{\cal L}^*$ during their accretion phases.

There may be another way to produce BSSs.
A nascent rapid rotating magnetized NS could form with a mass
reaching the Oppenheimer limit, but quickly losses its angular
momentum via gravitational (driving the rotation-modes unstable)
and electromagnetic (magnetic dipole) radiations.
A NS central density has to be high enough for a phase conversion
to a strange star\footnote{This process was supposed to be a
candidate of the ``center engines'' of $\gamma$-ray bursts (e.g.,
astro-ph/9908262).} before the NS is so slow and cool that a
super-Oppenheimer mass is possible.
Such a strange star should also be bare since 1, the phase
transition energy $\gg$ the crust gravitational binding; 2, the
photon emission rate $\gg L_{\rm Edd}$.

In conclusion BSSs can exist in nature. Probably some of them may
act as those X-ray sources in Table 1.

\section{Cooling and thermal emission of bare strange stars}

We could expect a nascent strange star with thermal energy ${\cal
E}_{\rm i} \ga 10^{52}$ ergs since the gravitational and the
degenerate energies are in the same order, $\sim 10^{53}$ ergs,
even if other energy sources (e.g., the rotation energy, the phase
transition energy) are included.
The specific heat of strange quark matter is (e.g., Usov 2001b)
%\begin{equation}
$C=C_{\rm q}+C_{\rm e}$,
%\label{C}
%\end{equation}
with $C_{\rm q}=1.9\times 10^{12}\rho_{15}^{2/3}T_{\rm
c}\exp(-\Delta/T)[2.5-1.7T/T_{\rm c}+3.6(T/T_{\rm c})^2]$ ergs
cm$^{-3}$K$^{-1}$, $C_{\rm e}=1.3\times 10^{11}Y_{\rm
e}^{2/3}\rho_{15}^{2/3}T$ ergs cm$^{-3}$K$^{-1}$.
The specific heat of unpaired electrons dominates, $C_{\rm
e}>C_{\rm q}$, when $T<7.45 \times 10^9$ K.
The electron fraction $Y_{\rm e}\sim 10^{-3}$. The energy gap is
very uncertain, and whether the color super-conducting (CSC)
occurs is therefore still a question. We choose $\Delta=50$ MeV
for next discussion, so a strange star should be in CSC state
except for the very beginning of its birth. The critical
temperature $T_{\rm c}\sim \Delta/2$ in the BCS model.
By ${\cal E}_{\rm i}=CT_{\rm i}\cdot 4\pi R^3/3$, one obtains the
initial temperature $T_{\rm i}\ga 10^{10}$ K, which means strange
is very hot soon after supernova explosion.

Effective neutrino emissivity of a newborn hot strange star
rapidly expels the thermal energy, making the strange star have a
much cooler temperature at which the photon emission dominates.
%(Schaab et al. 1996)
%
The dividing temperature $T_\nu$ is a solution of
\begin{equation}
3\times 10^{-4}\sigma T_\nu^4=R\epsilon_\nu(T_\nu), \label{Tnu}
\end{equation}
which is $T_\nu\sim 4\times 10^{10}$ K for typical parameters,
where the neutrino emissivity (e.g., Usov 2001b)
%\begin{equation}
$\epsilon_\nu=7.8\times 10^{-28}\alpha_{\rm s}Y_{\rm
e}^{1/3}\rho_{15}T^6\exp(-\Delta/T)~{\rm ergs~cm^{-3}~s^{-1}}$,
%\label{enu}
%\end{equation}
and $\sigma$ is the Stefan-Boltzmann constant. The factor
$10^{-4}$ in Eq.(\ref{Tnu}) is due to the upper limit on photon
emissivity of strange quark matter at energies $<20$ MeV (Chmaj et
al. 1991). This $T_\nu$ estimated is on the high side if CSC does
not occur at the very beginning, nevertheless this value implies
photon emission almost dominates all over a strange star's life.
This conclusion is strengthened if the Usov's (1998) photon
emission mechanism is included.

The equation governing a BSS's cooling history is
\begin{equation}
{4\over 3}\pi R^3\cdot C{{\rm d}T\over {\rm d}t}=-\xi\sigma
T^4\cdot 4\pi R^2, \label{Tg}
\end{equation}
where $\xi\sim 1$ for $T>10^9$ K at which the Usov mechanism
works, whereas $\xi\la 10^{-4}$ for $T<8 \times 10^8$ K (Usov
2001c).
When $T<7.45 \times 10^9$ K, assuming a constant $\xi$,
Eq.(\ref{Tg}) can be solved to be
\begin{equation}
T=T_0(1+2{\cal J}\xi T_0^2t)^{-1/2}, \label{T}
\end{equation}
where ${\cal J}=3\sigma/({\check C}_{\rm e}R)$, ${\check C}_{\rm
e}=C_{\rm e}/T$, $t$ is the time duration when a BSS cools from
temperature $T_0$ to $T$.
According to Eq.(\ref{T}), a BSS cools to $T\sim 10^6$ K after
$\sim 10^3$ years.
However, because of the magnetospheric polar cap heating, powered
by the bombardment of downward-flowing particles, a BSS should
keep a minimum temperature $T_{\rm min}$. As a rough estimate at
first, equating the photon emission rate to the pulsar spindown
power, $\xi\sigma T_{\rm min}^4\cdot 4\pi R^2\sim 6.2\times
10^{27}B_{12}^2R_6^6(2\pi/P)^4$, one has
%\begin{equation}
$T_{\rm min}=3.4\times 10^6R_6B_{12}^{1/2}P^{-1}~{\rm K}$.
%\label{Tmin}
%\end{equation}
For PSR J0437-4715 and PSR B0833-45, $T_{\rm min}\sim 9\times
10^2R_6$ eV and $6\times 10^3R_6$ eV, respectively. Considering
that the photon emission power is $\xi\la 10^{-4}$ times that of a
blackbody, these temperatures, modified by a factor of $\sim 0.1$,
are comparable with observations (Table 1).
%
%Nonetheless, the cooling calculations presented above are rough
%for these two special radio pulsars.

In fact the minimum temperature is model-dependent, and it is
worth to discuss $T_{\rm min}$ in some pulsar emission models.
%although no certain model can explain all of the observations.
Due to the high binding energy of bare quark surface, the
space-charge-limited flow model (e.g., Arons \& Scharlemenn 1979)
can not work for BSSs. We focus thus on the polar cap heatings in
the vacuum polar model (Ruderman \& Sutherland 1975) and the outer
gap model (Cheng, Ho \& Ruderman 1986), both of which are depicted
in Xu et al. (2001). The polar heating rate of RS-type gap is
$\sim 1.1\times 10^{31}\gamma_7 B_{12} P^{-2}$ ergs/s,
and the minimum temperature is thus
%\begin{equation}
$T_{\rm min}^{\rm RS}\simeq 3.5\times 10^6\gamma_7^{1/4}
R_6^{-1/2}B_{12}^{1/4}P^{-1/2}~{\rm K}$,
%
%\label{TminRS}
%\end{equation}
with $\gamma=10^7\gamma_7$ the typical Lorentz factor of the
primary particles. If outer gaps exist, the luminosity deposited
onto the surface is
$\sim 8.2\times 10^{30} B_{12} P^{-5/3}$ ergs/s,
and the correspondence temperature is
%\begin{equation}
$T_{\rm min}^{\rm CHR}\simeq 3.3\times 10^6
R_6^{-1/2}B_{12}^{1/4}P^{-5/12}~{\rm K}$.
%
%\label{TminCHR}
%\end{equation}
We see that these three values of $T_{\rm min}$, $T_{\rm min}^{\rm
RS}$, and $T_{\rm min}^{\rm CHR}$ are almost the same. This is not
surprising because, although the total energy deposit fluxes
differ, the thermal temperature is the amount flux to the power of
1/4.

As for the AXPs (or SGRs), the thermal energy with temperature
$T_{\rm min}$ can not account for their persistent X-ray
emissivity since the observed X-ray power is many orders higher
than the energy loss rate of rotarion. Nonetheless there are
actually two suggestions for extra energy supplying:
magnetism-powered (the so called magnetar model, e.g., Thompson \&
Duncan 1995) and accretion-powered (e.g., Marsden \& White 2001).
Both these mechanisms have been widely discussed in literatures
recently.
Since BSSs can also act as magnetars as long as the dynamo action
in proto-strange stars is effective enough (see Xu \& Busse 2001),
we deem that magnetic field line reconnection on BSS surfaces can
also work to produce abundant energy.
As discussed in section 2 (see eq.{\ref{L*}}), accretion in AXPs
and SGRs can still keep a strange star to be bare. So it is also
possible that AXPs (or SGRs) are accretion powered BSSs. Therefore
the energy budget problem of AXPs and SGRs are solved if those two
popularly discussed mechanisms are adapted to fit BSSs.

In principle, one can study the thermal radiative properties by
comparison of theoretically modelled spectra with that of
observations.
Unfortunately no emergent spectrum calculation of BSSs appears in
literature. The total power of photon emissivity of BSSs was done
by Chmaj et al. (1991). Nevertheless we could expect that the
spectra could be close to blackbodies, which represents the
general apparent of the X-ray spectra observed, since, e.g., for
the quark bremsstrahlung radiation mechanism (Chmaj et al. 1991),
quarks are nearly in thermal equilibrium by inter-collisions
within a depth less than the mean free path ($\sim 10$ fm) of
photons with energy $<\hbar \omega_{\rm p}\sim 20$ MeV.
A BSS with surface temperature $T$ may have a slightly harder
spectrum than that of a blackbody with $\sim 10^{-1}T$.
New fits by BSS emergent spectra may alter significantly the
physical quantities derived through the thermal radiations. For
example, one powerlaw and only one thermal spectra might be enough
to model precisely the observed spectrum of PSR J0437-4715 (Zavlin
et al. 2002).
Because of this lack of fits, the temperatures and radii listed in
Table 1 may not be relevant if we want to obtain observationally
the thermal properties (e.g., the temperature distribution) on a
BSS surface.

This kind of research may get a real information of photons from
quark matter astrophysically, whereas in terrestrial physics,
direct photons and lepton pairs have been recognized to be the
clearest signatures for quark-gluon plasma (e.g., Cassing \&
Bratkovskaya 1999). It is worth noting that the BSS thermal photon
emission is in the low energy limit, which would thus complement
the study of the high energy photons of relativistic
nucleus-nucleus collisions.
Future observations in various ways may confirm the existence of
BSSs, and in return the observational fit of the thermal spectra
from the quark surfaces could be used as a test in checking those
phenomenological models for quark gluon plasma in strong magnetic
fields.

It should be noted that magnetospheric power law components of
BSSs are also featureless (Xu \& Qiao 1998, Xu et al. 2001), but a
neutron star may have magnetospheric line features because of the
ions, pulled out from NS surface by the space-charge-limited-flow
mechanism, in the open field line region.

%\section{X-ray transients: bare strange stars?}

\section{Conclusion \& discussion}

An alternative opinion is proposed for the nature of the sources
with featureless X-ray spectra observed by Chandra and XMM, which
is that these X-ray emitters are simply bare strange stars (BSSs).
Possible scenarios to create a BSS are studied, and we find that
accretion can not prevent from forming a BSS unless the accretion
rate is much higher than the Eddinton one. The cooling and the
thermal radiation of a BSS are also discussed, indicating that
they are not strongly conflict with observations.

There could be indications for one or two lines at about 40\AA~
and 20\AA~ in RX J1856.5-3754 (van Kerkwijk 2002). This is a real
challenge for the BSS idea. If future longer observations with
Chandra and XMM confirm the existence, the source is certainly not
a BSS, but could be a crusted strange star since stringent
constrains on the mass ($M\approx 1 M_\odot$) and radius
($R\approx 6$ km) for RX J1856.5-3754 (Ransom et al. 2001) show
clearly that it can hardly be modelled by the equations of states
of nuclear matter.

The age of PSR J0437-4517 is worth deliberating. A millisecond
pulsar could be very hot soon after recycling phase when the polar
heat is transported effectively to the other part of BSS due to a
small $\varepsilon$. However, r-mode instability may spin down a
BSS to an initial period $P_0\sim 3-5$ ms (e.g., Andersson \&
Kokkotas 2001), which can have substantial influence on the age
calculation by dipolar radiation braking. However $P_0$ is
temperature dependent, and thus relevant to the accretion history.
The fastest rotating pulsar, PSR 1939+21, might have a small
accretion rate but a long time during its accretion phase. The age
of PSR J0437-4517 is much smaller if its $P_0\sim 5$ ms soon after
accretion, and it thus has high temperature today.

This is a critical time in obtaining the thermal spectra from
pulsar-like compact stars.
Besides Chandra and XMM, more X-ray missions (HETE-II in 2003,
Astro-E in 2005) may finally reveal the secrets, including whether
some of the sources are BSSs. We are looking forward to the
discoveries over the coming years.

%\vspace{0.2cm} %
%\noindent {\it Acknowledgments}:
%
%This work is supported by National Nature Sciences Foundation of
%China (10173002), and by the Special Funds for Major State Basic
%Research Projects of China.

\section*{Appendix: Energy re-emitted on the polar caps}

The essential difference between the accretion- or
rotation-powered energy deposit processes of NSs and BSSs is that
part of the energy should be transported to the outsite of the
polar caps for BSSs, but not for NSs (e.g., Xu et al. 2001), since
the coefficient of thermal conductivity of electron in the neutron
star surface
%\begin{equation}
$\kappa^{\rm NS} = 3.8\times 10^{14} \rho_5^{4/3} ~{\rm ergs}~
{\rm s}^{-1}~ {\rm cm}^{-1}~ {\rm K}^{-1}$,
%\label{kappaNS}
%\end{equation}
%
is much smaller than the coefficient of degenerate quark matter
%\begin{equation}
$\kappa^{\rm BSS}=\kappa^{\rm BSS}_{\rm q}+\kappa^{\rm BSS}_{\rm
e}$,
%\label{kappabss}
%\end{equation}
with $\kappa^{\rm BSS}_{\rm q} = 1.41\times 10^{21}\alpha_{\rm
s}^{-1} \rho_{15}^{2/3} \exp(-\Delta/T)~ ~{\rm ergs}~ {\rm
s}^{-1}~ {\rm cm}^{-1}~ {\rm K}^{-1}$ for quark scattering,
$\kappa^{\rm BSS}_{\rm e} = 1.55\times 10^{23}Y_{\rm e
}\rho_{15}T_9^{-1}~ ~{\rm ergs}~ {\rm s}^{-1}~ {\rm cm}^{-1}~ {\rm
K}^{-1}$ for electron scattering (e.g., Blaschke et al. 2001),
$\rho_5$ and $\rho_{15}$ being the densities in unit of $10^5$ and
$10^{15}$ g cm$^{-3}$, respectively, $\alpha_{\rm s}$ the coupling
constant of strong interaction, $T$ the temperature,
$T_9=T/(10^9{\rm K})$, the energy gap $\Delta\sim 10-100$ MeV, and
$Y_{\rm e}\sim 10^{-3}$ the ratio of numbers of electrons and
baryons.
A dimensional argument gives out the temperature difference
between polar cap and equator for BSSs,
\begin{equation}
\delta T \sim {\cal L}/(\kappa^{\rm BSS}R), \label{delT}
\end{equation}
where ${\cal L}$ is the rate of total energy deposit, $R\sim 10^6$
cm is the stellar radius.
For ${\cal L}\sim 10^{36}$ ergs s$^{-1}$, one has $\delta T\sim
6.5\times 10^9 T_9$ K, which is in the same order of polar
temperature.\footnote{%
In case of no energy dissipated (e.g., for NSs), the polar
temperature $T=2.3\times 10^6{\cal L}_{30}^{1/4}P^{1/4}$, where
${\cal L}_{30}={\cal L}/(10^{30}{\rm ergs/s})$. This temperature
is an upper limit of that of BSSs.
} %
This means substantial energy should be dissipate to outside of
polar cap in BSS if ${\cal L}\la 10^{36}$ ergs s$^{-1}$.
Defining $\varepsilon$ to be the re-emission photon fraction of
the total energy deposit, we have $1>\varepsilon>r_{\rm
p}^2/(2R^2)\sim 10^{-4}/P$ if ${\cal L}\la 10^{36}$ ergs s$^{-1}$,
where $r_{\rm p}=1.45 \times 10^4P^{-1/2}$ cm is the polar cap
radius, $P$ the rotation period.
Assuming one-half of the deposit energy being brought away by
neutrinos rather than photons in this case, one obtains modified
limits for $\varepsilon$: $5\times 10^{-5}/P<\varepsilon<0.5$,
where the upper limit is for ${\cal L}\rightarrow +\infty$ and the
lower limit for ${\cal L}\rightarrow 0$.

%
% -------------------- Table 1 -----------------------
\begin{deluxetable}{lcccccc}
\tablewidth{7in} \tablenum{1}%
\tablecaption{``Neutron stars''\tablenotemark{a}~ with X-ray
thermal spectrum
 observed by Chandra or XMM}%
\tablehead{ \colhead{Name} & \colhead{Period} & \colhead{B-Field
(G)} & \colhead{Temperature (eV)\tablenotemark{b}} &
\colhead{Radius (km)\tablenotemark{b}} & \colhead{$\gamma_{\rm
pl}$\tablenotemark{c}} & \colhead{Age (y)} }
\startdata%
RX J1856.5-3754 (INS) & - & - & 20(g), $\sim 60$(l) &
$\leq 10$(g), 2.2(l) & - & $\sim 10^6$ \nl%
RX J0720.4-3125 (INS) & 8.39 {\rm s} & - & 86(l) &
 - & - & - \nl%
1E 1048.1-5937 (AXP) & 6.45 {\rm s} & Magnetar? & $\sim 600$(l)
& - & $\sim 3$ & - \nl%
4U 0142+61 (AXP) & 8.69 {\rm s} & Magnetar? & 418(l) &
- & 3.3 & - \nl%
PSR J0437-4715 (msPSR)\tablenotemark{d} & 5.76 {\rm ms} & $3\times
10^8$ &
181(core), 46.5(rim) & 0.1(core), 2(rim) & 2.2 & $4.9\times 10^6$(?) \nl%
PSR B0833-45 (Vela) & 89.3 {\rm ms} & $3.4\times 10^{12}$ & 129(l)
& 2.1(l) & 2.7 & $1.1\times 10^4$ \nl%
PSR B0656+14 (PSR) & 385 {\rm ms} & $4.7\times 10^{12}$ &
69.0(g), 138(l) & 22.5(g), 1.7(l) & - & $1.0\times 10^5$ \nl%
% \hline
\enddata
\tablenotetext{a}{References: Burwitz et al. (2001), Pons et al.
(2002), Paerels et al. (2001), Tiengo (2002), Juett et al. (2002),
Zavlin et al. (2002), Sanwal et al. (2002),  Marshall \& Schulz
(2002).}
\tablenotetext{b}{``g'' and ``l'' denote for {\it g}lobal and {\it
l}ocal (e.g., polar-cap) blackbody spectra, respectively.}
\tablenotetext{c}{The photon index of a nonthermal power-law
spectrum.}
\tablenotetext{d}{The temperature and radius here are for the
fitting of the data with the two-temperature hydrogen polar-caps,
but could be qualitatively similar parameters for a
two-temperature blackbody polar-cap model.}
\label{t.a}
\end{deluxetable}{}

\end{document}